\documentclass[aps, preprint]{revtex4}
\usepackage{amsfonts}
\usepackage{amsmath}
\usepackage{amssymb}
\usepackage{graphicx}

\begin{document}

\title{Perfect coupling of light to surface plasmons with ultra-narrow
linewidths}
\author{M. Sukharev$^{1\ast }$, P. R. Sievert$^{2}$, T. Seideman$^{3}$, and
J. B. Ketterson$^{2,4}$}
\affiliation{$^{1}$Department of Applied Sciences and Mathematics, Arizona State
University at the Polytechnic Campus, Mesa AZ, 85212, USA\\
$^{2}$Department of Physics and Astronomy, Northwestern University, Evanston
IL, 60208, USA\\
$^{3}$Department of Chemistry, Northwestern University, Evanston IL, 60208,
USA\\
$^{4}$Department of Electrical and Computer Engineering, Northwestern
University, Evanston IL, 60208, USA\\
$^{\ast }$corresponding author: maxim.sukharev@asu.edu}

\begin{abstract}
We examine the coupling of electromagnetic waves incident normal to a thin
silver film that forms an oscillatory grating embedded between two otherwise
uniform, semi-infinite half spaces. Two grating structures are considered,
in one of which the mid point of the Ag film remains fixed whereas the
thickness varies sinusoidally, while in the other the mid point oscillates
sinusoidally whereas the film thicknesses remains fixed. On reducing the
light wavelength from the long wavelength limit, we encounter signatures in
the transmission, $T$, and reflection, $R$, coefficients associated with: i)
the short-range surface plasmon mode, ii) the long-range surface plasmon
mode, and iii) electromagnetic diffraction tangent to the grating. The first
two features can be regarded as generalized (plasmon) Wood's anomalies
whereas the third is the first-order conventional (electromagnetic) Wood's
anomaly. The energy density at the film surface is enhanced for wavelengths
corresponding to these three anomalies, particularly for the long range
plasmon mode in thin films. When exciting the silver film with a pair of
waves incident from opposite directions, we find that by adjusting the
grating oscillation amplitude and fixing the relative phase of the incoming
waves to be even or odd, $T+R$ can be made to vanish for one or the other of
the plasmon modes; this corresponds to perfect coupling (impedance matching
in the language of electrical engineering) between the incoming light and
these modes.
\end{abstract}

\maketitle

\section{Introduction}

\label{introduction}There is currently much interest in the physical
properties and possible applications of plasmons excited in various metallic
structures \cite%
{AtwaterJAppPhys2005,OzbayScience2006,AbajoRMP2007,BarnesAdvMat2007,CatrysseNatPhys2007,Lezec98,Kawata08}%
, particularly those made from silver or gold, where relatively narrow
resonances are observed relative to some other metals \cite{ZhangJPCC2008}.
Much of this interest is associated with the high electric fields generated
by various resonant responses. Common among the many applications of these
fields is the enhancement of nonlinear optical responses for optical devices
\cite{FendlerAdvMat2004} or spectroscopic applications \cite%
{VanDuyneExpRevMolD2004}. The simplest plasmonic structure is a single
metallic sphere. In the limit where the sphere diameter is small compared to
the wavelength of the incident light, the polarization is well described by
the leading term in the Mie expansion, corresponding to a single peak at $%
\omega =\omega _{bp}/\sqrt{3}$, where $\omega _{bp}$ is the bulk plasmon
frequency \cite{KreibigBook}. As the sphere size increases higher order
terms in the Mie expansion become important \cite{Mie}.

Smaller sphere radii yield larger field enhancements at the particle
surface. For ellipsoids \cite{Nitzan82}, when the exciting field lies along
the major axis of a prolate ellipsoid of revolution, the high curvature for
the surface normal to that direction further enhances the field; this is the
so-called lightning rod effect. One is then encouraged to examine structures
with sharp corners, such as pyramids \cite{SchatzPyramids}, or assemblies of
these containing adjacent sharp corners leading to higher local field
enhancements. Although the local fields associated with small finite
structures and their arrays can be relatively high, they are in practice
limited by various effects including boundary scattering and radiation
losses. With regard to the latter, one is encouraged to examine structures
where radiation is suppressed.

The specific case of surface plasmons has long been studied \cite%
{RaetherBook}. When propagating at a half space separating a bulk metal from
an adjacent dielectric, radiation is kinematically forbidden; damping only
involves scattering from inhomogeneities (e.g. a rough surface) and
dissipation in the metal itself. In fact the very existence of a surface
plasmon requires that its frequency, $\omega _{p}$, and wavevector, $k_{p}$,
not match light propagating at any angle in the surrounding dielectric. For
this reason it is common to couple to surface plasmons via an evanescent
light wave and carefully control the coupling strength. In the commonly used
Otto geometry \cite{OttoOriginal1968} this is accomplished by forming a thin
dielectric layer between a prism and the free surface of the metal adjacent
to it. By adjustment of the thickness of the coupling layer, the reflected
beam in the prism can be made to vanish at some critical angle \cite%
{QueizerATR1973}, a condition we will refer to as \emph{critical coupling}.

If, in the Otto geometry, the metal layer (generally a vapor deposited thin
film) has a finite thickness, plasmons involving both sides can be excited.
If the dielectric constant of the coupling layer and that of the far side
differ, modes with different wavelengths for a given frequency are present,
which tend to localize on opposing sides of the metal film. When the
dielectric constants on both sides are identical, however, the modes become
degenerate; if in addition the film is thin, the modes are strongly coupled
and split, with one being symmetric and the other antisymmetric with respect
to the mid point of the film \cite{RaetherBook}. The structure of these
modes was first examined by Economou \cite{Economou69}. Later work by Sarid
\cite{SaridPaper} showed that the damping of the symmetric modes is greatly
reduced as the film thins leading to long range propagation; the
antisymmetric mode turns out to have a much shorter range.

This paper focuses on coupling to the long- and short-range modes using a
sinusoidal grating for the case of perpendicular incidence involving a
special case of Wood's anomaly, which we discuss at greater length below.
Grating coupling has been used to experimentally couple to plasmon modes at
a free surface, in connection with a study of the associated one-dimensional
plasmon bandstructure \cite{Seymore83}. More recently the plasmon
bandstructure for thin square-wave modulated films has been examined \cite%
{Kawata08}. Structures similar to those modeled here have recently been
studied experimentally and will be reported elsewhere \cite{Ketterson09}. On
the practical side, when coupled with an appropriate dye, the structures may
be used to make future vertically emitting plasmonic band-gap lasers \cite%
{Kawata04}.

The paper is arranged as follows. In Section \ref{Sarid geometry} we
consider electromagnetic modes in the Otto geometry in symmetric dielectric
environments modeled with a dielectric constant, $\varepsilon _{Ag}\left(
\omega \right) $ corresponding to silver embedded in a water environment. We
also present calculations of intensity enhancements and line widths for
critical coupling. Section \ref{FDTD} describes the methodology applied in
our numerical simulations and illustrates results for in-phase and
out-of-phase oscillatory gratings using the same silver and dielectric
environment parameters used for the Otto geometry simulations. Specifically,
we calculate surface averaged field enhancements, demonstrating large
enhancements at the Wood's anomalies, in particular for the long-range mode.
Our conclusions are summarized in Section \ref{Discussion and Conclusion}.

\section{Long- and short-range plasmons in thin, flat metal films}

\label{Sarid geometry}Although the magnetic field amplitude is largest
within the metal for the symmetric mode, the corresponding electric field
has a node at the mid point and its over-all amplitude is smaller than for
the antisymmetric mode; hence it has a smaller dissipation, leading to a
longer range (and correspondingly a longer lifetime). One of our goals in
the present study is to achieve the largest possible electric field
enhancements in a spatially resonant, film-based structure and therefore low
dissipation is a desirable feature. We will refer to this symmetric mode as
the \emph{long-range plasmon}. In addition the energy density lies largely
outside the film, with the result that the mode velocity approaches that of
light in the surrounding dielectric as the film is thinned \cite{RaetherBook}%
; the mode can be thought of as being \textquotedblleft
stiff\textquotedblright\ in this sense.

On the other hand the electric field for the antisymmetric mode is,
over-all, larger within the metal. This leads to greater damping as the film
is thinned; simultaneously the velocity falls (being \textquotedblleft
loaded\textquotedblright\ by the metal) and the mode is \textquotedblleft
soft\textquotedblright\ in this sense. We therefore refer to this mode as
the \emph{short-range plasmon}.

In modeling the overall behavior of the incoming beam in an Otto geometry
experiment, one can regard the system as a stack of dielectric layers and
apply the Fresnel conditions at all interfaces, while simultaneously
accounting for the propagation between interfaces. In the present study, we
consider a silver film and approximate the dielectric function by the Drude
model,
\begin{equation}
\varepsilon _{1}(\omega )=\varepsilon _{\infty }-\frac{\omega _{bp}^{2}}{%
(\omega ^{2}+i\omega \Gamma )}  \label{Drude}
\end{equation}%
where the parameters are $\omega _{bp}=1.7901\times 10^{16}$ rad/sec, $%
\Gamma =3.0841\times 10^{14}$ rad/sec, and $\varepsilon _{\infty }=8.926$.
This model is a good representation of silver films for wavelengths between $%
300$ nm and $700$ nm. We define $k_{0}=2\pi /\lambda _{0}=\omega /c$, the
wave vector of the exciting light of wavelength $\lambda _{0}$ in vacuum,
where $c$ is the velocity of light. For very thin films the effects of
scattering from an additional boundary would enter; no attempt to
incorporate such effects is made in what follows. Note the effects of
scattering from a single boundary are implicitly contained in the measured
reflectivity data from which the dielectric constant is obtained.

Using dispersion relations for the plasmon modes in the Sarid geometry and
our model parameters one can show that the high (long-range) and low
(short-range) frequency modes lie, respectively, above and below that for a
surface plasmon propagating at a half space, where the latter is given by
\begin{equation}
\frac{\omega }{\omega _{bp}}=\frac{1}{\sqrt{\varepsilon _{\infty
}+\varepsilon _{2}}},  \label{ratio}
\end{equation}%
here we will take the dielectric constant of the host medium, $\varepsilon
_{2}$, is $1.76$, corresponding to water. Eq. (\ref{ratio}) results in $0.306
$ for our model parameters. Both families of curves approach this limit as
the film thicknesses increases. This indicates the validity of our solution,
although it is not strictly within the range of applicability of the Drude
model. We also note that as the real in-plane plasma momentum decreases, all
curves merge with the "light-line" for the dielectric medium, which is given
by $(\omega /\omega _{bp})=(1/\sqrt{\varepsilon _{2}})(k_{\rho }^{\prime
}/k_{bp})$; here $k_{bp}$ is the in-plane wave vector of the bulk plasmon
and $k_{\rho }^{\prime }$\ is the real part of the in-plane complex wave
vector of the plasmon. As it was found in \cite{SaridPaper}, the high
frequency mode has a very long range for thin films. As noted above, this is
due to destructive interference of the overlapping evanescent fields in the
metal, which minimizes the energy dissipation for this mode and enhances the
lifetime of the plasmons at the surface. The dissipation decreases
quadratically with the film thickness and linearly with the imaginary part
of $\varepsilon _{1}$.

\begin{figure}[tbph]
\centering\includegraphics[width=\linewidth]{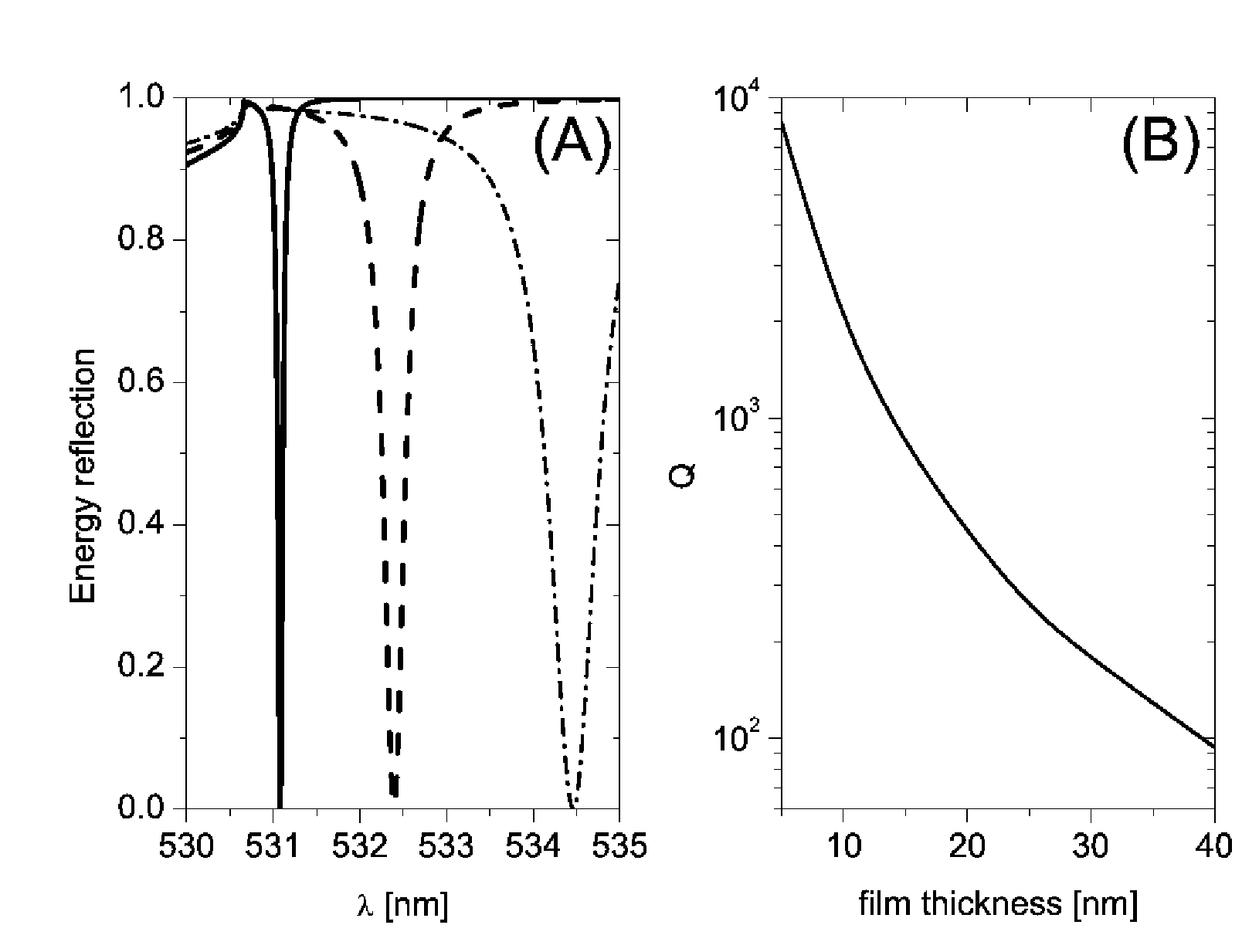}
\caption{(A) Energy reflection coefficient vs wavelength at critical
coupling for film thicknesses of $5$ nm (solid curve), $10$ nm (dashed
curve), and $15$ nm (dash-dotted curve). (B) Quality factor, $Q=\protect%
\lambda /\Delta \protect\lambda $, for the surface plasmon resonance vs
silver film thickness derived from the linewidths shown in Fig. \protect\ref%
{Fig1}A.}
\label{Fig1}
\end{figure}

For the purpose of our paper it is desirable to couple into the long-range
plasmon mode with the minimum radiation loss, as this should maximize the
field strength on the metal film. For a featureless, flat, metal film and
fixed excitation frequency, the only method available is "tuning", i.e.
varying the thickness, of the dielectric (coupling) layer between the
coupling prism and the metal film. The total reflection coefficient is then
calculated from the generalized Fresnel coefficients, accounting for the
multiple reflections from the layers involved. The incident light wavelength
is swept while the nominal angle of incidence is varied to correspond to a
fixed in-plane momentum $k_{\rho }^{\prime }$ of $(2\pi /400)$ nm$^{-1}$
(this fixed value of $k_{\rho }^{\prime }$ is arbitrary but corresponds to
the periodicity of the grating couplers considered in Section \ref{FDTD}
below). Excitation of the plasmon modes is indicated by minima in the total
reflection. The minimum corresponding to either mode can be "tuned" to zero
by adjusting the thickness of the dielectric layer. This is the critical
coupling layer thickness and corresponds to perfect coupling or impedance
matching. The energy reflection at critical coupling for the long-lived mode
versus the wavelength for a number of silver film thicknesses is given in
Fig. \ref{Fig1}A. One can determine a linewidth at the half-power line at
various film thicknesses and calculate the $Q$-factor, defined as $Q=\lambda
/\Delta \lambda $. The results are shown in Fig. \ref{Fig1}B. Note that the $%
Q$ significantly increases as the silver layer thins and increases by $4$
orders of magnitude at a thickness of $5$ nm, implying a corresponding build
up of the stored energy; the electric field, which scales as the square root
of the energy density, is similarly enhanced.

\section{Optics of metal gratings at impedance matching conditions}

\label{FDTD}
\begin{figure}[tbph]
\centering\includegraphics[width=\linewidth]{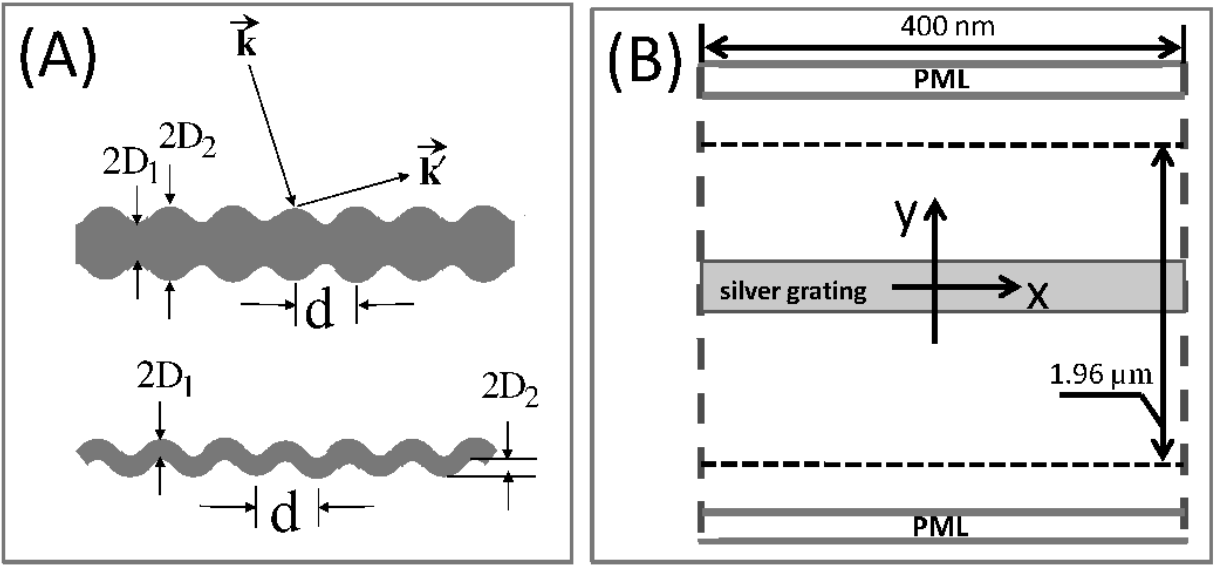}
\caption{(A) Schematic setup of out-of-phase and in-phase sinusoidal
gratings. (B) Unit cell of FDTD simulations. The vertical dashed lines
depict the periodic boundaries. Two horizontal dashed lines represent the
detection contours. The metal film shown in the center of the unit cell is
excited either by a laser pulse generated along a single horizontal line
placed a few spatial steps beneath the upper PML region (we refer to this
excitation scheme as a single-ended excitation), or by two incident plane
waves generated symmetrically on both sides of the film with a fixed phase
delay between the incident pulses (referred to as double-ended excitation
scheme). }
\label{Fig2}
\end{figure}

An alternative way of coupling into surface plasmons is to
arrange for the metal film to act as a diffraction grating, a "line grating"
in the present study. The choice of the grating structure introduces a wide
range of physical scenarios. A relatively simple choice involves a
sinusoidal variation of the position of the metal/dielectric interfaces and
here we will study two limiting cases the two limiting cases shown in Figs. %
\ref{Fig2}A, B. In Fig. \ref{Fig2}A the mid point of the metal film remains
fixed but the thickness varies sinusoidally between values $2D_{1}$ and $%
2D_{2}$ \cite{KorovinOptLett08}; in Fig. \ref{Fig2}B we have a case where
the mid point oscillates sinusoidally with amplitude $2D_{2}$ but the film
thicknesses remains fixed at a value $2D_{1}$. The two sinusoids are,
respectively, out-of-phase and in-phase for these two cases, as shown in
Fig. \ref{Fig2}A. Intuitively we expect the modes for the constant thickness
case to closely resemble the Sarid modes on a flat film while simultaneously
allowing grating coupling to an external light wave.

For an incoming direction and wavelength for which the grating diffracts the
light tangent to the grating, such that the diffracted wave is
\textquotedblleft in step\textquotedblright\ with the grating itself, one
expects a stronger coupling; this enhanced interaction manifests itself as
anomalies in other diffracted orders including the zeroth-order transmitted,
$T$, and reflected, $R$, light intensity. Note that when energy is absorbed
by the film, the condition $T+R=1$, which one would have for a stack of
lossless dielectric layers, does not apply. The phenomena, first discovered
by R. W. Wood, is referred to as \emph{Wood's anomaly}; the theoretical
explanation was first given by Rayleigh and for this reason it is also
called the Rayleigh-Wood anomaly. For a rigorous discussion the reader is
referred to reference \cite{Oliner65}.

\begin{figure}[tbph]
\centering\includegraphics[width=\linewidth]{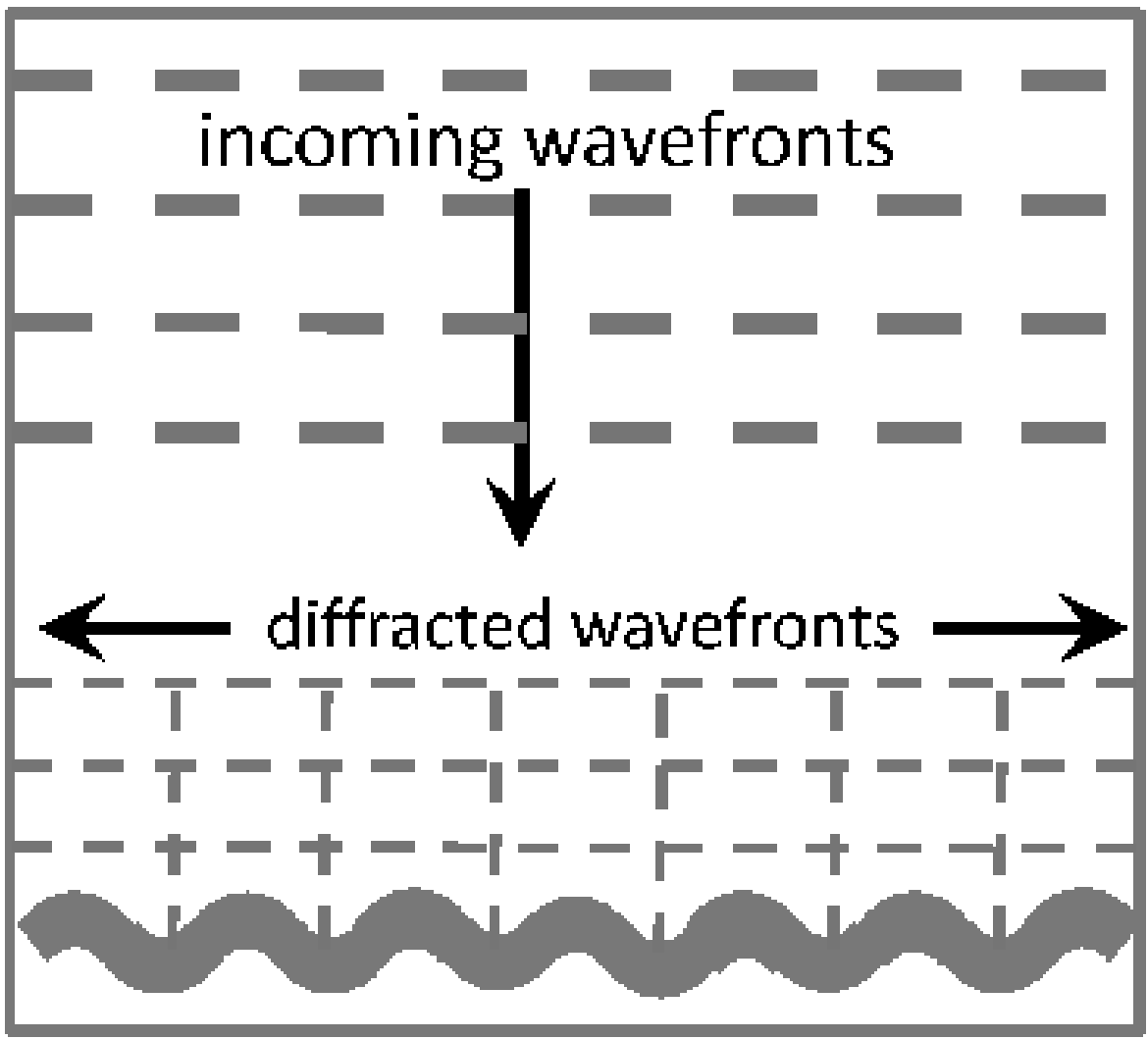}
\caption{The electromagnetic Wood's phenomenon in which an incoming wave
(entering perpendicular) is diffracted tangent to the grating such that the
wave fronts match the grating spacing resulting in anomalies in the
reflected and transmitted waves.}
\label{Fig3}
\end{figure}

The Wood's phenomena is shown schematically in Fig. \ref{Fig3} for the case
of a wave entering perpendicular to the grating. In this case we have equal
and opposite diffracted waves (for a total of two, unlike the case for an
arbitrary incident angle) that form a standing wave, and one may anticipate
an even stronger interaction.

As a generalization of the Rayleigh-Wood phenomena, we will include the two
cases where the wavelength of the long- and short-range plasmons (rather
than the electromagnetic wave) matches the grating spacing, which we can
interpret as \emph{plasmon Wood's anomalies}. We note in passing that when
the incoming wave is perpendicular to the film one couples to the plasmons
at the second Brillouin zone point of the associated one-dimensional
plasmonic band structure within the film that is associated with the
presence of the periodic grating.

The optical properties of the periodic metal structures depicted in the Fig. %
\ref{Fig2} were simulated using FDTD \cite{TafloveBook} in two dimensions.
We have restricted ourselves to the case of normal incidence; furthermore we
limit the calculations to the transverse magnetic mode, TM, which couples to
plasmons propagating in the same direction as the static grating waves. For
two dimensions this mode is sometimes designated as the transverse electric
mode \cite{TafloveBook}, TE$_{z}$, and has two electric and one magnetic
component. The form given in Eq. (\ref{Drude}), with identical parameters,
forms the basis for the dielectric response of the metal to external EM
excitation.

The FDTD\ unit cell used in the simulations is shown schematically in the
Fig. \ref{Fig2}B. The upper and lower parts of the grid are terminated by
horizontal absorbing boundaries, for which we use perfectly matched layers
(PML) \cite{Berenger} with a depth of $16$ spatial steps and exponential
differences in order to avoid diffusion instabilities. In all simulations we
use a uniform Yee grid with spatial steps $\delta x=\delta y=0.98$ nm and a
time step $\delta t=\delta x/\left( 1.5c\right) $. The metal film is assumed
to be embedded in a dielectric with refractive index of $1.33$,
corresponding to $\epsilon _{2}=1.76$. We note that the latter leads to a
Wood's anomaly at $532$ nm for the $400$ nm periodic gratings as discussed
in the previous section. The gratings are excited by $x$-polarized plane
waves that are generated along horizontal lines near the PML\ regions. All
simulations have been performed in parallel using $128$ processors on the
DataStar cluster at the San Diego Supercomputer Center and Saguaro cluster
at the ASU High Performance Computing Center. An average execution time of
our codes is approximately $20$ minutes or less.

Since the FDTD algorithm is a time domain integrator, electromagnetic (EM)
field components at some pre-defined incident wavelength can be obtained
either by applying a long excitation pulse, with duration much longer than
the characteristic resonant lifetime, or by exciting the system with an
ultra-short incident pulse and subsequently Fourier transforming the
resulting EM fields. The latter method requires a single FDTD run within
which one obtains the entire spectrum. Here we employ simulations of phasor
functions using a discrete Fourier transform "on-the-fly" procedure with a
short pulse excitation scheme \cite{TafloveBook,SukharevPRB07}. In order to
calculate the transmission, $T$, and reflection, $R$, coefficients, we
numerically integrate the Poynting vector normal to the horizontal lines
shown in Fig. \ref{Fig2}B as dashed lines. Transmission, $T$, is calculated
using the total fields, which consists of the EM waves scattered by the
metal film and the incident field, whereas reflection, $R$, is simulated
using the scattered fields only, so as to exclude standing waves generated
by interference between the incident field and the EM waves scattered back
by the film. Numerical convergence is verified by ascertaining that $T+R=1$
for a lossless dielectric for all wavelengths of interest. For the
double-ended excitation scheme, i.e. for the case of excitation of the
grating from both sides of the film, we perform simulations of the
reflection for both sides of the structure using scattered fields only.
Finally, in order to assure numerical convergence especially in the vicinity
of sharp resonances we integrate Maxwell equations for as long as $2$ ps.

Unlike the Otto geometry discussed above, we do not in general expect to be
able to achieve critical coupling for our gratings, in the sense that all of
the incoming energy is absorbed by the film; this would require two
quantities to simultaneously vanish, namely the transmission and reflection
coefficients, whereas we have only a single parameter, the amplitude of the
grating oscillation, to vary (here we assume the thickness, or average
thickness for the out-of-phase gratings, is held constant). For this reason
we performed simulations where the beams simultaneously enter normal to the
grating but from opposite sides. On symmetry grounds we anticipated that if
these waves were in phase or out of phase we might selectively excite either
the short-range or long-range plasmon modes, but not both; this was
confirmed by the simulations, although the couplings for in-phase and
out-of-phase sinusoids were reversed as will be discussed below. For such a
case we then have only a single quantity, $T+R$, which has the same
magnitude on both sides and simulations show that we may cause it to vanish
(corresponding to critical coupling) at some critical value of our single
grating amplitude parameter. We note that if the oscillation amplitudes on
the upper and lower surfaces were independently varied (while maintaining
constant average thickness), critical coupling should be achievable under
single ended excitation, since we then have two independent parameters with
which to simultaneously tune $T$ and $R$ through zero.

\begin{figure}[tbph]
\centering\includegraphics[width=\linewidth]{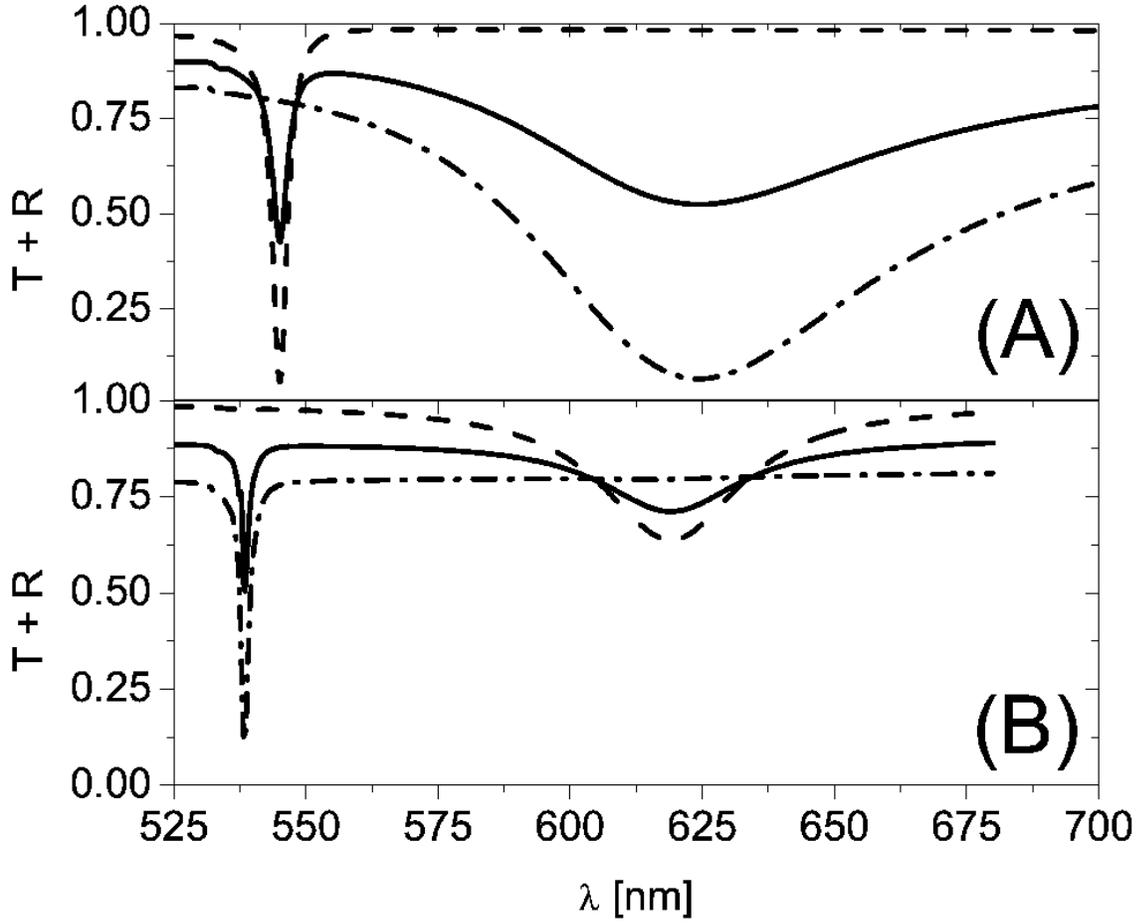}
\caption{Sum of the transmission, $T$, and reflection, $R$, coefficients as
a function of the incident wavelength. The solid curve illustrates the
single-ended excitation scheme, whereas the dashed and dash-dotted curves
show the double-ended excitation with phase shifts $\protect\pi $ and $0$.
Panel A corresponds to the out-phase sinusoidal grating with $D_{1}=5$ nm
and $D_{2}=31$ nm. Panel B shows data for the in-phase sinusoidal grating
with parameters $D_{1}=10$ nm and $D_{2}=14.4$ nm. The structural parameters
for both gratings have been optimized so as to minimize $T+R$ for the double
ended excitation scheme and the long-range mode. (Panel A - dashed curve,
panel B - dash-dotted curve).}
\label{Fig4}
\end{figure}

Fig. \ref{Fig4}A shows simulations for an out-of-phase grating (see Fig. \ref%
{Fig2}A) and calculate $T+R$ as a function of the free space incident
wavelength, $\lambda _{0}$, for three excitation schemes: i) single-ended
excitation (solid line); ii) symmetric double-ended excitation (dash-dotted
line) (where two in-phase plane waves emanating from symmetrically placed
horizontal lines near the upper and lower PML regions excite the metal); and
iii) anti-symmetric double-ended excitation (dashed line) (which is similar
to the symmetric scheme except that the phase difference between the two
incident waves is $\pi $). The single-ended excitation scheme clearly
generates all three modes: an electromagnetic Wood's anomaly located at $%
\lambda =532$ nm and the long-range and short-range surface plasmon modes
localized at $545$ nm and $624$ nm, respectively. We note that the resonant
wavelength for the electromagnetic Wood's anomaly accurately corresponds to
the grating period of $400$ nm in the dielectric media with the refractive
index of $1.33$.

It is apparent in Fig. \ref{Fig4}A that the phase delay between the two
incident pulses in the double-ended excitation scheme gives us a control
over the symmetry of the excited surface plasmon mode. The observed behavior
can be summarized as follows: for the case of out-of-phase sinusoidal
gratings (which are symmetric with respect to the film center), in-phase
incoming waves excite the symmetric mode whereas out-of-phase waves excite
the antisymmetric mode. On the other hand, Fig. \ref{Fig4}B shows
calculations for an in-phase sinusoidal grating where the converse is true:
in-phase incoming waves excite the antisymmetric mode whereas out-of-phase
waves excite the symmetric mode.

Provided that our gratings do not significantly depart from flat films, we
expect the anti-symmetric and symmetric resonant wavelengths to be
approximately given by the long- and short-range modes, which are governed
by the film thickness as given in Fig. \ref{Fig1}. For the in-phase gratings
the thickness is given by $2D_{1}$ and we can then use the parameter $2D_{2}$
to control the coupling. For the out-of-phase sinusoids the resonant
wavelength would, in the first approximation, be governed by the average
thickness, $\left( D_{1}+D_{2}\right) $; the difference, $\left(
D_{1}-D_{2}\right) $, can then be adjusted to control the coupling.

\begin{figure}[tbph]
\centering\includegraphics[width=\linewidth]{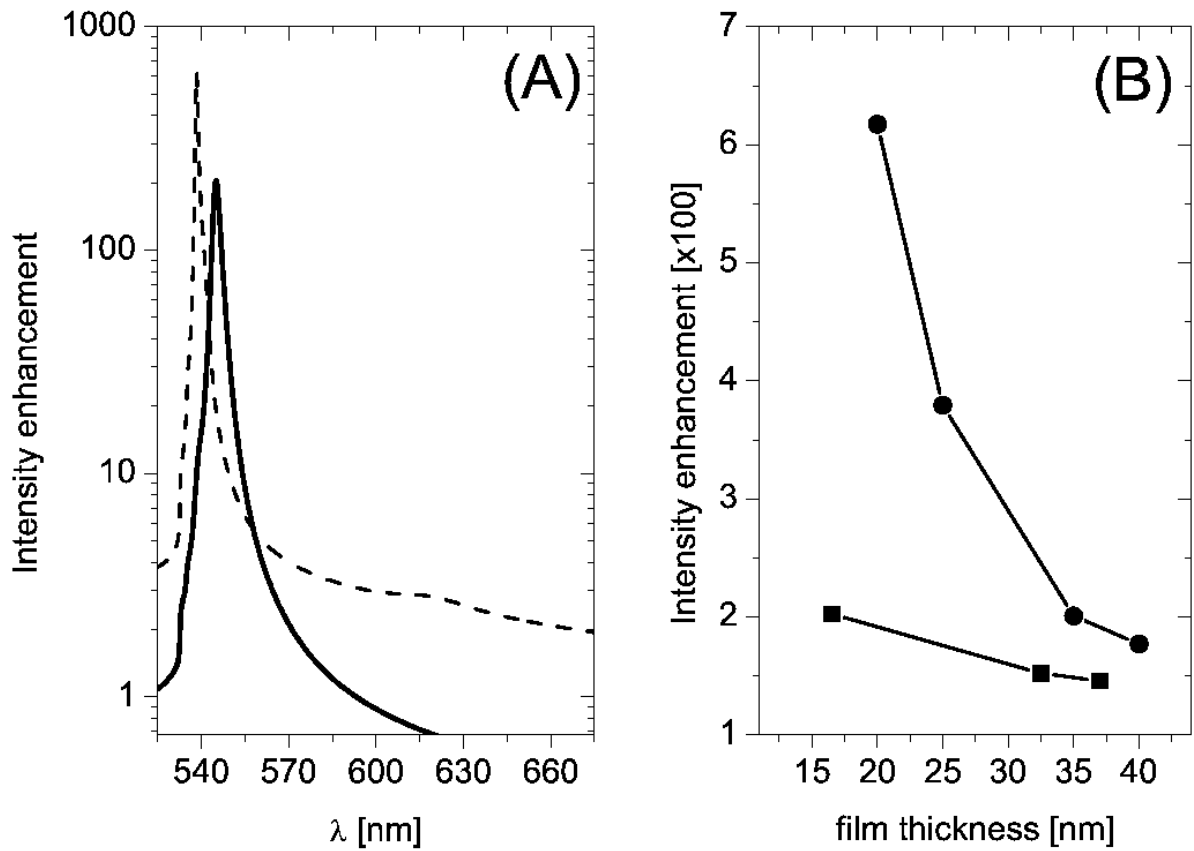}
\caption{(A) Surface averaged intensity enhancement as a function of the
incident wavelength for out-of-phase (solid curve) and in-phase (dashed
curve) gratings at the impedance matching conditions for the long-range
plasmon mode. (B) Surface averaged intensity enhancement as a function of
the film thickness at the impedance matching conditions for long-range
plasmon modes for the out-of-phase (squares) and in-phase (circles)
sinusoidal grating.}
\label{Fig5}
\end{figure}

Fig. \ref{Fig4}A shows an example of an out-of-phase sinusoidal grating
described by the parameters $D_{1}=5$ nm and $D_{2}=31$ nm. Here we achieve
a near perfect coupling for both the short- and long-range modes with the
same parameter set. Note that with the single ended excitation the maximum
coupling is approximately $50\%$. Fig. \ref{Fig4}B shows an example of an
in-phase grating where the parameters $D_{1}=10$ nm and $D_{2}=14.4$ nm
minimize $T+R$ for the long-range surface plasmon mode. As discussed in
previous sections, at the impedance matching conditions incident EM
radiation is efficiently transformed into propagating surface plasmons
leading to strong field enhancements on the surface of the gratings. Fig. %
\ref{Fig5}A presents the intensity enhancement factors averaged over the
metal surface for the out-of-phase (solid line) and in-phase (dashed line)
sinusoidal grating under the double-ended excitation conditions (here we
consider only long-range modes). We note that high EM fields, significantly
enhanced at the impedance matching conditions, are not localized near
specific spatial regions, as in the case of tip-enhanced optical probes \cite%
{TERS}, as expected fort a standing sinusoidal wave. Following the results
of Section \ref{Sarid geometry} we simulated the enhancement for several
grating thicknesses as shown in Fig. \ref{Fig5}B. The results confirm our
general conclusions that with thinner films the intensity enhancements are
much higher as follows from the high quality factors $Q$ (see Fig. \ref{Fig1}%
B for details).

\section{Discussion and Conclusions}

\label{Discussion and Conclusion}The above calculations demonstrate that the
largest enhancements occur for the long-range mode for an in-phase
oscillation of the grating amplitude. In evaluating nonlinear optics
applications requiring rapid response, one must keep in mind that resonance
is always accompanied by dispersion and the associated distortion of
waveforms; the number of field cycles required to build up a response is
proportional to the $Q$ of the resonance and this may be particularly
important when, e.g., trying to exploit plasmon waveguiding. Hence we
envision the large enhancements obtained here would primarily be exploited
for sensor applications.

We recall that the large intensity enhancements for our grating structures
given in section \ref{FDTD} were calculated on a per-unit-area basis. This
property is particularly important when considering sensor applications
involving fluorescence-based spectroscopy or Surface Enhanced Raman
Spectroscopy (SERS) where we assume we have no control over where the target
molecule will actually attach. It is sometimes assumed that the enhancement
of SERS scales as the square of the intensity; although this appears
physically unreasonable, we note that the surface averaged square of the
intensity is also greatly enhanced. Although various plasmon resonant
nanostructures can have \textquotedblleft hot spots\textquotedblright\ with
rather large intensity enhancements, exploiting these requires that the
target molecules \textquotedblleft find\textquotedblright\ these hot spots,
which in turn requires effective mixing or long exposure times. If receptors
are involved one would want them attached only on the hot spots; if they are
widely distributed they will compete with the hot spots for target molecules
thereby depleting the latter.

Many other one-dimensional grating structures come to mind, including
grooves and slots; both have multiple parameters (thickness, width, vertical
position) associated with them, which can be tuned to achieve critical
coupling under single ended excitation. The sharp edges associated with
these structures can lead to further, but highly local, enhancements.
Two-dimensional (e.g. hole) arrays are natural extensions. Our group is
currently examining several of these structures. For SERS, and especially
for fluorescence, the response of the grating at the emission (shifted)
wavelength is also important, an issue under study.

In connection with possible SERS applications we point out that it should be
possible to engineer one- and two-dimensional structures where light is i)
gathered and ii) concentrated in a \textquotedblleft two
step\textquotedblright\ process involving a periodic array of bumps, ridges
or holes. Firstly, the periodic grating structure so formed would be
engineered to efficiently collect the light in an impedance matched manner,
as we have described. Second, the edges of the structures so patterned would
optimize to concentrate these fields.

We now discuss several technical issues regarding the implementation of the
structures we have modeled. We used water for the dielectric in our
simulations; i.e. we are implicitly assuming water-based sensor
applications. In order to support the film, we require an insulating coating
on a substrate whose dielectric constant can be adjusted to equal that of
water. Such a material is Teflon AF\copyright\ manufactured by Dupont.
Although forming a sinusoidal grating is challenging, one can pattern and
etch a grooved structure in, e.g., a glass or Si substrate by a variety of
techniques, where the groove edges naturally develop a slope under simple
etching protocols. Subsequently spinning on of Teflon AF would, via surface
tension, further smooth the contour, thereby better approximating a sine
wave. Depositing a uniform Ag film on a low surface tension material like
Teflon AF may also be problematic. We note that silver reacts with water
over time and so one might use Au (for which the enhancements will be lower)
or spin on a very thin protective layer of Teflon AF over or within which
any receptors would be incorporated \cite{Nenninger}.

Finally we must address the issue of matching the grating resonance
frequency for the long-range mode to that of an available laser, say an
argon ($517$ nm) or a YAG ($532$ nm) laser. Forming the structure with the
required $d$-spacing is not an option, given the narrowness of the
resonances. One can, however, take advantage of the degree of freedom
mentioned in the Introduction, namely tipping the incoming laser beam out of
the scattering plane, in particular in a direction parallel to the grating
lines (corresponding to $\theta =\pi /2$). By making the $d$-spacing
slightly smaller than that which would match the resonance one can tune
through the resonance by decreasing $\theta $ slightly from the plane
normal. At resonance we would retain a standing wave along the grating plane
and directed normal to its lines but the plasmon wavevector would then have
a small component parallel to the lines.

\begin{acknowledgments}
This research was supported in part by the NCLT program of the National
Science Foundation at the Materials Research Institute of Northwestern
University (grant number ESI-0426328) and the AFOSR\ DARPA program (grant
number FA9550-08-1-0221). The numerical work was enabled by the resources of
the National Science Foundation San Diego Supercomputer Center under grant
number MCA06N016.
\end{acknowledgments}

\newpage

\newpage

\newpage

\newpage

\end{document}